\documentclass[12pt]{article}

\usepackage{amssymb}

\def\a{{\alpha}}
\def\b{{\beta}}

\def\d{{\delta}}
\def\l{{\lambda}}
\def\m{{\mu}}

\newcommand{\ee}{\mbox{\bf e}_-}
\newcommand{\ef}{\mbox{\bf e}_+}
\newcommand{\eh}{\mbox{\bf h}}
\def\bD{{\bf \Delta}}

\newcommand{\bs}{\mbox{\bf S}}
\newcommand{\sm}{\mbox{\bf s}}

\def\on#1#2{\mathop{\vbox{\ialign{##\crcr\noalign{\kern2pt}
$\scriptstyle{#2}$\crcr\noalign{\kern2pt\nointerlineskip}
\kern-2pt$\hfil\displaystyle{#1}\hfil$\crcr}}}\limits}

\def\nn{ \nonumber }
\def\bq{ \begin{equation} }
\def\eq{ \end{equation} }
\def\ben{ \begin{eqnarray} }
\def\en{ \end{eqnarray} }
\def\frac#1#2{{#1\over #2}}
\def\dfrac#1#2{{\displaystyle{#1\over#2}}}

\newtheorem{prop}{Proposition}

\begin{document}
%%%%%%%%%%%%     TITLE   %%%%%%%%%%%%%%
\title{Canonical transformations of the extended phase space,
Toda lattices and St\"{a}ckel family of integrable systems. }
\author{
A.V. Tsiganov\\
\it\small Department of Mathematical and Computational Physics,
 Institute of Physics,\\
\it\small  St.Petersburg University,
 198 904, St.Petersburg, Russia.\\
\it\small e-mail: tsiganov@mph.phys.spbu.ru}
 \date{}
\maketitle

%%%%%%%%     A B S T R A C T     %%%%%%%%%

\begin{abstract}
We consider compositions of the transformations of the time variable
and canonical transformations of the other coordinates, which map
completely integrable system into other completely integrable system.
Change of the time gives rise to transformations of the integrals of
motion and the Lax pairs, transformations of the corresponding
spectral curves and $R$-matrices. As an example, we consider
canonical transformations of the extended phase space for the Toda
lattices and the St\"ackel systems.
\end{abstract}
\vfill
\newpage

%%%%%%%%%%%%  TEXT  %%%%%%%%%%%%%%
\section{Introduction}
It is well known, in classical mechanics any canonical transformation
of variables maps a given integrable system into other integrable
system. In this paper we consider compositions of the change of the
time variable $t$ and canonical transformations of the other
coordinates, which map completely integrable system into other
completely integrable system.

On the $2n$-dimensional symplectic manifold $\cal M$ (phase space)
with coordinates $\{p_j,q_j\}_{j=1}^n$ let us consider hamiltonian
system determined by the Hamilton function $H(p,q)$. By definition
canonical transformation of the phase space $\cal M$ have to preserve
canonical form of the Hamilton equations
\bq
\dot{q}_i=\dfrac{\partial{ H(p,q)}}{\partial p_i}\,,\qquad
\dot{p}_i=-\dfrac{\partial{ H(p,q)}}{\partial q_i}\,.\qquad
\label{hameq0}
\eq
As sequence, the action integral
\[S=\int\left(\sum_{i=1}^n p_i\,dq_i -H\,dt\right)\,,\]
differential form $\sum_{i=1}^n p_i\,dq_i$ on $\cal M$
 and the Poisson brackets on $\cal M$ are invariant
with respect to the canonical transformations of the phase space
$\cal M$.

At the Hamilton equations (\ref{hameq0}) and at the canonical
transformations of $\cal M$ the time $t$ plays a role of parameter.
If we want to consider change of the time $t$, we have to add new
coordinate $q_{n+1}=t$ with the corresponding momenta $p_{n+1}=H$ to
the phase space $\cal M$ \cite{lanc49}. The resulting
$2n+2$-dimensional space ${\cal M}_E$ \cite{lanc49,syng60} is
so-called extended phase space of the hamiltonian system.

Canonical functional $S$ on ${\cal M}_E$ has the following completely
symmetric form
\[S=\int_{\tau_1}^{\tau_2}\sum_{i=1}^{n+1} p_i\,q'_i\,d\tau\,.\]
On the extended phase space ${\cal M}_E$ the Jacobi, Euler-Lagrange
and Hamilton variational principles $\delta S=0$ are differed by an
additional constraint
\bq
{\cal H}(p_1,\ldots,p_{n+1};q_1\,\ldots,q_{n+1})=0\,.\label{oham}
\eq
Here $\cal H$ is called generalised Hamilton function \cite{lanc49}.
According by (\ref{oham}), any hamiltonian system is a conservative
system on the extended phase space ${\cal M}_E$.

As an example, the Hamilton principle with generalised hamiltonian
\[{\cal H}=p_{n+1}-E=H-E\]
gives rise to the Hamilton equations on the extended phase space
${\cal M}_E$
\bq
\dot{q}_i=\dfrac{\partial{\cal H}}{\partial p_i}\,,\qquad
\dot{p}_i=-\dfrac{\partial{\cal H}}{\partial q_i}\,,\qquad
i=1,\ldots,n+1\,.\label{hameq}
\eq
On zero-valued energy surface  ${\cal H}=0$ these equations are the
initial Hamilton equations (\ref{hameq0}) on the phase space ${\cal
M}$  and two additional equations
\[\dot{t}=1\,,\qquad \dot{E}=\dfrac{\partial{H}}{\partial t}\,.\]
Below we shall consider conservative hamiltonian systems at $\partial
H/\partial t=0$ only.

By definition canonical transformations of the extended phase space
${\cal M}_E$ preserve canonical form of the Hamilton equations
(\ref{hameq}) or the Hamilton-Jacobi equation
\bq
\dfrac{\partial S}{\partial t}+H=0\,.\label{hjeq}
\eq
Invariance of these equations leads to the invariance of the
canonical differential form $\sum_{i=1}^{n+1} p_i\,dq_i$ with respect
to canonical transformations of the space ${\cal M}_E$
\cite{lanc49}. Thus, change of the time variable
\bq t\mapsto\widetilde{t}\,,\qquad d\widetilde{t}=v(p,q)\,dt
\label{ttr}
\eq
and transformation of the corresponding momenta $H$
\bq
H\quad\mapsto\quad
\widetilde{H}=\dfrac{H}{v(p,q)}\,
\label{tr1}
\eq
are canonical transformation of the extended phase space. Here $H$
and $\widetilde{H}$ are variables of the extended phase space, but
simultaneously they may be considered as the functions on the initial
phase space \cite{lanc49}. In general relativity function $v(p,q)$
(\ref{ttr}) is so-called lapse function, which determines
transformation from physical time to coordinate time
\cite{mis73}.

Canonical transformations of the extended phase space ${\cal M}_E$
may be defined for any hamiltonian system. However, among all the
hamiltonian systems a set of the completely integrable hamiltonian
systems attracts a special attention. It is known, this set is
invariant by canonical transformations of the initial phase space
$\cal M$. In this paper, we want to discuss the following
\par\noindent
{\bf Problem:~} {\it How to construct canonical transformation of the
extended phase space ${\cal M}_E$, which maps a given integrable
system into the other integrable system.}

An exact technical definition of a completely integrable system is
provided by the well-known Liouville theorem. Briefly, such system
possesses $n$ functionally independent integrals $\{I_j\}_{j=1}^n$ of
motion in the involution
\[\{I_j,I_k\}=0\,,\qquad j,k=1\,\ldots,n\,,\qquad H=I_1\,.\]
Thus, to construct canonical transformation of ${\cal M}_E$
preserving integrability, we have to explicitly determine new algebra
$\widetilde{\cal A}_I$ of integrals by using a given algebra ${\cal
A}_I$ of integrals of motion.

For an integrable system we can consider dynamics with respect to the
time $t_j$ conjugated to the integral $I_j$. Thus, in generic, we can
add to the phase space $\cal M$ all the integrals of motion $\{I_j\}$
with the conjugated times $\{t_j\}$. The geometry of the
$4n$-dimensional extended phase space ${\cal M}_I={\cal
M}\oplus_j\{I_j,t_j\}$ may be used to study of the transformations of
the integrals of motion and conjugated to them times $t_j$.

By the Liouville theorem we can introduce the action-angle variables
$s_1,\ldots,s_n$ and $\varphi_1,\ldots,\varphi_n$ in the vicinity of
the common level surface of integrals
\[
{\cal M}_\a=\left\{z\in {\cal M}:~I_j(z)=\a_j\,,~j=1,\ldots,n\,.
\right\}
\]
Transformation of the time $t$ (\ref{ttr}) acts on the half of the
equations of motion only
\[
\dfrac{d\varphi_j}{dt}=\omega_j(s_1,\ldots,s_n)
\quad\mapsto\quad
\dfrac{d\varphi_j}{d\widetilde{t}}
=\dfrac{\omega_j(s_1,\ldots,s_n)}{v(t,s,\varphi)}
\,,
\]
whereas other equations are invariant
\[\dfrac{ds_j}{dt}=0 \quad\mapsto\quad  \dfrac{ds_j}{d\widetilde{t}}=0\,.\]
If the lapse function $v\Bigl(\,p(s,\varphi),q(s,\varphi)\,\Bigr)$
depends on the action variables only
\[v(s,\varphi)=v(s),\]
change of the time (\ref{ttr}) preserves canonical form of the
equations of motion in the action-angle variables
\bq
\dfrac{ds_j}{d\widetilde{t}}=0\,,\qquad
\dfrac{d\varphi_j}{d\widetilde{t}}=\widetilde{\omega}_j(s_1,\ldots,s_n)=
\dfrac{\omega_j(s_1,\ldots,s_n)}{v(s_1,\ldots,s_n)}
\,.\label{ctr}
\eq
Moreover, at $v=v(s)$ canonical transformations (\ref{ttr}-\ref{tr1})
of the extended phase space ${\cal M}_E$ map a given integrable
system into other integrable systems. In this case mapping
(\ref{tr1}) determines factorization of the initial Hamiltonian
\bq
H(s)= v(s)\cdot\widetilde{H}(s)
\label{exph}
\eq
up to canonical transformations of the action-angles variables
$\{s,\varphi\}$.

Thus, by the Liouville theorem we could construct canonical
transformations of the extended phase space, which map a given
integrable system into other integrable systems. However, it may of
course be quite difficult to construct the action-angles variables
for a given mechanical system, even if it is known to be completely
integrable. Below we shall construct canonical transformation
(\ref{ttr}-\ref{tr1}) in the physical variables only.

Change of the time preserving integrability induces transformation of
all the machinery developed for integrable systems. Let us recall
that the key idea which has started the modern age in the study of
classical integrable systems is to bring them into Lax form
\[\{H,L\}=[L,M]\,,\] see reviews \cite{rs87,pe91}. The first matrix $L$
or, more precisely, the coefficients of its characteristic polynomial
$P(\l)=\det(L-\l)$ determine integrals of motion. Change of the time
gives rise to an algebraic transformation of the Lax matrices, a
geometric transformation of the corresponding algebraic curves and
transformations of the separation of variables methods.  By using all
the possible algebraic and geometric transformations, we can try to
construct canonical transformations of the extended phase space
preserving integrability.

Transformations of the time variable and extensions of the phase
space have been discussed many times in classical and quantum
mechanics. The wittingly incomplete list includes:
\par\noindent
- solution of the equations of motion, as example for the Kowalewski
top \cite{kol01};
\par\noindent
- the Kowalewski-Painlev\'{e} analysis
\cite{rgb89};
\par\noindent
- study of singularities of solutions equations of motion
\cite{mos70};
\par\noindent
- geometrical theory of integrable system
\cite{lanc49,syng60};
\par\noindent
- qualitative theory of dynamical systems \cite{bog80,shar66};
\par\noindent
- construction of the bi-Hamiltonian \cite{brw94} and quasi
bi-Hamiltonian systems \cite{tm98};
\par\noindent
-  the Birman-Schwinger formalism in quantum mechanics \cite{rids82}.

As a first example, we consider integrable system determined by a
natural Hamiltonian
\bq
H(p,q)=H_0+V=\sum g_{ij}\,p_ip_j+V(q_1,\ldots,q_n)\,,\label{js1}
\eq
where $H_0(p)$ and $V(q)$ are kinetic and potential parts of the
Hamilton function. For any given energy $E$ of the system we can use
the Hamiltonian ${\cal H}=H-E$ to represent dynamics, provided that
we impose the constraint ${\cal H}=0$ \cite{lanc49,syng60}.  The
passage to the geometric representation is accomplished by the Jacobi
time transformation at
\[v(q)=E-V(q)\] such that
\[ dt=\dfrac{d\widetilde{t}}{E-V(q)}\,,\]
which maps orbits of an energy surface ${\cal H}=0$ into geodesics of
the Jacobi geometry, i.e. into the orbits of the following
Hamiltonian
\bq
\widetilde{H}=\dfrac{H_0}{E-V(q)}=\sum \widetilde{g}_{ij}\,p_ip_j\,,
\qquad \widetilde{g}_{ij}=(E-V)g_{ij}\,.\label{js2}
\eq
The new Hamiltonian $\widetilde{H}$ will then give the same equations
of motion on the surface $\widetilde{\cal H}=1$. Here for a
positive-defined kinetic energy $H_0$ the term $(E-V)$ is always
non-negative in the physically allowed region.  A modern discussion
of this canonical time transformation may found in
\cite{rosp95}.

Of course, the geodesic motion may be replaced by other systems. For
instance, let us consider two-dimensional oscillator
\[H=H_0+a\,V(q)=(p_1^2+p_2^2+b) +a\,(q_1^2+q_2^2)\,.\]
In this case free Hamiltonian $H_0(p,q)$ does not pure kinetic part
of $H$ (\ref{js1}). This choice may be considered as a shift of the
energy surface ${\cal H}=-b$. However, it is more convenient to
consider the free Hamiltonian $H_0$ as a sum of the kinetic energy
and the constant potential $U(q)=b$. For this system the Kepler
canonical transformation of the time variable (\ref{tr1}) with the
function
\[v(p,q)=V(q)=q_1^2+q_2^2\]
preserves integrability. This change of the time $t$ has been known
by Kepler, the corresponding transformation of the Hamilton function
has been studied by Levi-Civita \cite{lc06} and extended in
\cite{mos70}.

After change of the time (\ref{tr1}) and the point canonical
transformation of the other variables $(p_1,q_1,p_2,q_2)\to
(p_x,x,p_y,y) $ the orbits of the oscillator maps into the orbits of
the Kepler problem
\bq
\widetilde{H}=\dfrac{H_0(p,q)}{v(q)}+a=p_x^2+p_y^2+
\dfrac{b}{\sqrt{x^2+y^2}}+a\,.\label{keptr}
\eq
Both these systems are degenerate St\"ackel systems in plane. We can
construct at least four different Lax matrices for them.  According
to \cite{ts98b}, the Kepler change of the time induces the following
transformation of the Lax matrices
\bq
L(\l)\mapsto \widetilde{L}(\l)=L(\l)+ \left(\begin{array}{cc}
  0 & 0 \\ \widetilde{H} & 0 \end{array}\right)\,.
  \label{kepl}
\eq
The Lax matrix $L(\l)$ associated with the oscillator is defined on
the initial phase space $\cal M$. The Lax matrix $\widetilde{L}(\l)$
associated with the Kepler problem is defined on the extended phase
space ${\cal M}_E$. This transformation may be considered as a shift
of the initial Lax matrix $L(\l)$ on the matrix from the extended
phase space ${\cal M}_E$. Additional term in (\ref{kepl}) is the
constant matrix with respect to the new time.

Here both Hamiltonians $H$ and $\widetilde{H}$ linearly depend on
parameters $a$ and $b$. Let the origin integrable Hamiltonian
$H(p,q;\,a_1,\ldots,a_k)$ linearly depends on arbitrary parameters
$(a_1,\ldots,a_k)$. Then the function $v(p,q)$ (\ref{tr1}) may be
founded by using additional propositions, for instance, that other
integrals of motion are polynomials on parameters \cite{hgdr84}. In
this case the passage to the new time may be considered as a coupling
constant metamorphosis acting on the one of the constant $a_j$
\cite{hgdr84}. The similar additional propositions Kepler used by investigations of the celestial mechanics lows \cite{shar66}.

It is known, in quantum mechanics, the eigenvalue problem of the
Hamiltonian operator $H$
\[
H(p,q)\Psi=(H_0+a\,V+b)\Psi=E\Psi\,,
\]
may be joined with the eigenvalue problem of the charge operator $a$
\[
\widetilde{H}(p,q)\widetilde{\Psi}=
(\widetilde{H}_0+(b-E)\,
V^{-1})\widetilde{\Psi}=-a\widetilde{\Psi}\,.
\]
In quantum mechanics, such duality of the two eigenvalue problems has
been used by Fok \cite{fok35}, Schr\"odinger \cite{shr40} and many
other. In the Birman-Schwinger formalism function $v(q)$ is called a
"sandwich" potential
\cite{rids82}. The canonical transformation of the time variable
(\ref{tr1}) is an analog of this duality. Recall, in quantum
mechanics zeroes of the potential $v(p,q)=V(q)$ and signs of the
parameters $a$ and $b$ determine compactness or non-compactness of
the Sturm operator $\widetilde{H}$ in the corresponding space
\cite{rids82}. In the classical mechanics zeroes of the potential
$v(p,q)=V(q)$ and signs of the parameters $a$ and $b$ determine
behavior of the system with respect to the inversion of the time
\cite{shar66}.

Now let us consider an action of the canonical transformations of the
extended phase space on the algebra of integrals of motion. Let the
phase space of the initial system may be modelled on the dual space
of certain Lie algebra ${\mathfrak g}$. It is known, the set of
integrals $\{I_j\}$ gives rise to the subalgebra ${\cal A}_I$ in the
corresponding universal enveloping algebra $U({\mathfrak g}^*)$. A
classical system is called integrable if the commutant ${\cal A}_I$
of the Hamiltonian $H$ in $U({\mathfrak g}^*)$ contains an abelian
subalgebra of the necessary rank. Of course, different abelian
subalgebras in $U({\mathfrak g}^*)$ determine different integrable
systems on ${\mathfrak g}^*$. However, the modern representation
theory does not allows us to construct and classify all the possible
abelian subalgebras in $U({\mathfrak g}^*)$.

The search of the canonical transformations (\ref{tr1}) may be
reformulated as  the con\-struc\-tion of the abelian subalgebra in
$U({\mathfrak g}^*)$ by using information about the known algebra of
integrals ${\cal A}_I$. For this purpose algebra ${\cal A}_I$ is too
little, but algebra $U({\mathfrak g}^*)$ is too much. By using some
additional propositions, we want to extend the first algebra (or
restrict the second algebra).

Let the integrals of motions be either the fixed highest order
polynomials in momenta, either polynomials in parameters, such that
\bq
I_j=\sum_{i,k} I_j^{ik}\,p_k^i\,,\qquad\mbox{\rm ¨"¨}\qquad
I_j=\sum_{i,k}
\widehat{I}_j^{ik}\,a_k^i\,.\label{raz}
\eq
In the fist case instead of origin algebra ${\cal A}_I$ we can
introduce new algebra ${\cal A}_I^p
\supset {\cal A}_I$ generated by momenta and corresponding coefficients $I_j^{ik}$ in (\ref{raz}). In the second case new algebra ${\cal A}_I^a \supset {\cal A}_I $ is generated by parameters and coefficients $\widehat{I}_j^{ik}$. Expansions (\ref{raz}) give rise to grading of the new algebras ${\cal A}_I^{p,a}$ by power of momenta or parameters.

By the Liouville theorem initial system is integrable, if the
algebras ${\cal A}_I^{p,a}$ contain the abelian subalgebras of the
necessary rank. If one of the elements $v(p,q)=I_j^{ik}$ or
$v(p,q)=\widehat{I}_j^{ik}$ is invertible in a given representation,
we can introduce a completed algebra $\widetilde{\cal
A}_I^{p,a}=\{v^{-1},{\cal A}_I^{p,a}\}$, such that ${\rm
rank}\widetilde{\cal A}_I^{p,a}\geq{\rm rank}{\cal A}_I$. The search
of the two abelian subalgebras in the algebra $\widetilde{\cal
A}_I^{p,a}\}$ with elements $H$ and $\widetilde{H}$ (\ref{tr1}) is
equivalent to construction of the canonical transformation
(\ref{tr1}).

Next we can consider two integrable systems on a common phase space
$\cal M$. By using two algebras of integrals ${\cal A}_I$ and ${\cal
A}_J$ we can construct new algebra ${\cal A}_{IJ}={\cal
A}_I\oplus{\cal A}_J$ or ${\cal A}_{IJ}={\cal A}_I\otimes{\cal A}_J$.
Of course, algebra ${\cal A}_{IJ}$ should be much richer than ${\cal
A}_I$ and ${\cal A}_J$, separately. This algebra has two abelian
subalgebras, which associate with two initial integrable systems. If
one of the elements $v(p,q)=I_j$ or $v(p,q)=J_j$ is invertible, we
can introduce a completed algebra $\widetilde{\cal
A}_{IJ}=\{v^{-1},{\cal A}_{IJ}\}$ such that ${\rm
rank}\widetilde{\cal A}_{IJ}\geq{\rm rank}{\cal A}_I$. Now, existence
of the third abelian subalgebra in $\widetilde{\cal A}_{IJ}$ with
elements $\widetilde{H}$ (\ref{tr1}) is provided existence of the
canonical transformation (\ref{tr1}).

Obviously, initial algebras ${\cal A}_I$ and ${\cal A}_J$ contain all
the necessary information about the completed algebras
$\widetilde{\cal A}_{I}^{p,a}$ or $\widetilde{\cal A}_{IJ}$. Thus, we
have to extract this information only. As an example of the solution
of such algebraic problems, we consider integrable systems related by
canonical transformations with the Toda lattices and the St\"ackel
systems.

In the second Section we propose canonical change of the time for the
Toda lattices similar to the Kepler change of the time (\ref{keptr}).
Let us rewrite the Hamiltonian of the Toda lattice as a sum of the
free Hamiltonian $H_0$ and potential $V_\b$
\[H=H_0+a_\b\cdot V_\b=
\Bigl(\sum K(p,p) + \sum_{
\a\in P,~\a\neq\b}\,
 a_\a\,\cdot e^{\a(q)}\Bigr) + a_\b
e^{\b(q)} \,,\qquad a_\a,~a_\b\in{\mathbb R}\,,
\]
where $P$ is a system of simple roots and $\b$ is one of these roots.
Here free Hamiltonian $H_0$ is a sum of the kinetic part with the
nontrivial potential. Similar to the Kepler transformation
(\ref{keptr}), change of the time (\ref{ttr},\ref{tr1}) preserves
integrability at
\bq
v(q)=V_\b=e^{\b(q)}\,\qquad \b\in P\,.\label{ttrt}
\eq
The corresponding mapping of the Lax matrices looks like
(\ref{kepl}). To construct these new integrable systems we shall use
other automorphisms of the $A_1$ subalgebras in the grading simple
Lie algebra ${\mathfrak g}$ associated to the Toda lattice.

In the third Section we consider some pair of the St\"ackel systems
with a common St\"ackel matrix, defined by two families of integrals
$\{I_k\}_{k=1}^n$ and $\{J_k\}_{k=1}^n$ . In this case ratio of any
two integrals $\widetilde{H}=I_m/J_m$ (\ref{tr1}) and the inner
product $({\bf I}\wedge {\bf J})$ of the vectors ${\bf I}$ and ${\bf
J}$ on ${\mathbb R}^n$ may be used to determine of the third
integrable set of integrals of motion. To construct corresponding
change of the time (\ref{tr1}) we explicitly used the Jacobi ideas in
the separation of variables method. Recall, Jacobi proposed to
construct various classes of completely integrable systems starting
from a set of separated one-dimensional problems.

\section{The Toda lattices.}
\setcounter{equation}{0}
Before proceeding further, it is useful to recall some known facts
about generalised Toda lattices (all details may be founded in review
\cite{rs87}).

Let $\mathfrak g$ be a real, split, simple Lie algebra ${\rm
rank}\,{\mathfrak g}=n$. Let $K(,)$ be its Killing form, let
$\mathfrak a$ be a split Cartan subalgebra, $\Delta$ the associated
root system, $\Delta_+$ the set of positive roots and $P$ the system
of simple roots. For $\a\in\Delta_+$, let ${\mathfrak g}_\a$ be the
corresponding root space and $e_\a\in{\mathfrak g}_\a$ a root vector.
It will be convenient to normalize $e_\a$ in such a way that
$K(e_\a,e_{-\a})=1$.

The root space decomposition, ${\mathfrak g}={\mathfrak a}+\oplus_\a
{\mathfrak g}_\a$, gives rise a natural grading on $\mathfrak g$, the
so-called principal grading. Let ${\mathfrak g}_+=\oplus_{i\geq 0}
{\mathfrak g}_i$ be a Borel subalgebra of $\mathfrak g$ and
${\mathfrak g}_-$ be the opposite nilpotent subalgebra. The
${\mathfrak g}={\mathfrak g}_+ +{\mathfrak g}_-$ is the generalised
Gauss decomposition for simple Lie algebra, which induces a dual
decomposition  ${\mathfrak g}^*={\mathfrak g_+}^*+{\mathfrak g}_+^*$
and an injection $i={\mathfrak g}_+^*\hookrightarrow {\mathfrak
g}^*$. Let ${\cal O}_a\subset {\mathfrak g_+}^*$ denote the coajoint
orbit of some $a\in{\mathfrak g_+}^*$. The image $i({\cal O}_a)\in
{\mathfrak g}^*$ has, in general, nothing to do with the coajoint
orbit ${\cal O}_{i(a)}$ of $i(a)$ in ${\mathfrak g}^*$. So $i^*$ of
the Casimir function $\phi$ on ${\mathfrak g}^*$ can be a
complicated, non-constant function on ${\cal O}_a$. The
Adler-Kostant-Symes theorem claims that any two such functions
$i^*\phi_1$ and $i^*\phi_2$ are in involution with respect to the
natural symplectic structure on ${\cal O}_a$.

The Killing form $K$ allows us to identify ${\mathfrak g}^*_i$ with
${\mathfrak g}_{-i}$, so that
\[
{\mathfrak g}^*_+=\oplus_{i\leq 0}{\mathfrak g}_i\,,\qquad {\mathfrak
g}^*_-=\oplus_{i> 0}{\mathfrak g}_i\,.
\]
Choose a vector
\[a=\sum_{\a\in P}a_a\,e_{-\a}\,,\qquad a_\a\in{\mathbb R}\]
and let ${\cal O}_a$ be the ${\mathfrak g}_+$ orbit of $a$ in
${\mathfrak a}+{\mathfrak g}_{-1}$. The points of ${\cal O}_a$ have
the form
\[
\xi=p+\sum_{\a \in P}\,c_\a\,a_\a\,e_{-\a}\,,\qquad p\in\
{\mathfrak a}\,,\qquad c_\a>0.
\]
The orbit ${\cal O}_a$ does not really depends on the $a_\a$'s but
only on their signs.

It is convenient to introduce the variable $q\in {\mathfrak a}$ such
that $c_\a=\exp \a(q)$. Fix a basis $\{h_j\}$ in $\mathfrak a$, and
let $\{h'_j\}$ be the dual basis with respect to the Killing form
$K(h_i,h'_j)=\d_{ij}$. Set
\[
q_j=K(q,h_j)\,,\qquad p_j=K(p,h'_j)\,,\qquad p,q\in {\mathfrak
a}\,.\] Then
\[
\{p_i,p_j\}=\{q_i,q_j\}=0\,,\qquad\{p_i,q_j\}=\d_{ij}\,,\]
i.e. the variables $\{p_j,q_j\}$ are canonical. The orbit ${\cal
O}_a$ is parametrized by the canonical variables as follows
\[
\xi= \sum_{i=1}^n p_ih_i+\sum_{\a\in P} a_\a\cdot\exp\Bigl(\sum_{i=1}^n
q_iK(\a,h'_i)\,\Bigr)\cdot e_{-\a},.
\]
Replace the inclusion $i={\mathfrak g}_+^*\hookrightarrow {\mathfrak
g}^*$ by its shifted version $i+e$, where $e$ is an element of
${\mathfrak g}^*$ which annihilates $[{\mathfrak g}_\pm,{\mathfrak
g}_\pm]$.

For the Toda lattice the orbit ${\cal O}_a$ is an orbit of
${\mathfrak g}_+$, what we are really interested in the orbits of the
full algebra ${\mathfrak g}_R={\mathfrak g}_+\oplus{\mathfrak g}_-$.
Let us translate ${\cal O}_a$ by adding to it a constant vector
$e=\sum_{\a\in P} e_\a$ which may be regarded as a one-point orbit of
${\mathfrak g}_-$. The resulting orbit
\bq
{\cal O}_{ae}={\cal O}_a+e \label{orbt}
\eq
is parametrized by the canonical variables as follows
\bq
L= \sum_{i=1}^n p_ih_i+\sum_{\a\in P} a_\a\cdot\exp K(\a,q)\cdot
e_{-\a}+\sum_{\a\in P} e_\a\,.
\label{todal}
\eq
Let us consider the simplest Hamiltonian on ${\cal O}_{ae}$ generated
by the Killing form on $\mathfrak g$
\bq
H(X)=\dfrac12\,K(X,X)\,.
\label{caz1}
\eq
Its restriction to ${\cal O}_{ae}$ (\ref{todal}) is given by
\bq
H(p,q)=\dfrac12K(p,p)+ \sum_{\a\in P}\,a_\a\,e^{\a(q)}\,.
\label{todah}
\eq
To keep in mind consequent canonical transformation of the time, we
choose the second Lax matrix at the following special form
\bq
\dfrac{d}{dt}\,L=[L,L_{-}]\,,\qquad
L_-=-\sum_{\a\in P} a_\a\cdot\exp K(\a,q)\cdot
e_{-\a}\,.\label{todalax}
\eq
Any finite-dimensional linear representation of $\mathfrak g$ gives
rise to a matrix valued function $L$ (\ref{todal}) on $\cal M$, the
coefficients of its characteristic polynomial are integrals of
motion.

The behavior of the dynamical system with the Hamilton function
(\ref{todah}) depends crucially on the signs $a_\a$. The associated
hamiltonian flow is complete if and only if $a_\a\geq 0$. In this
case, its solution is reduced to the generalised Gauss decomposition
in the corresponding Lie group. Systems with other signs of $a_\a$
are of course also solvable, but their solutions blow up \cite{rs87}.

Now let us construct canonical change of the time variable for these
Toda lattices. Recall, in a shifted version of the
Adler-Kostant-Symes scheme in order to get orbit ${\cal O}_{ae}$ in
${\cal M}\simeq{\mathfrak g}_R^*={\mathfrak g}_+^*\oplus{\mathfrak
g}_-^*$ we translate orbit ${\cal O}_a$ living in ${\mathfrak g}_+^*$
by adding to it a constant vector $e$ from the remaining part of
$\cal M$. Let us replace the phase space $\cal M$ on the extended
phase space ${\cal M}_E$. In this case we can also translate the same
orbit ${\cal O}_{a}$ in ${\mathfrak g}_+^*$ by adding to it a
constant vector from the remaining part of the whole space ${\cal
M}_E$. As above, this vector has to be a character and a constant
with respect to the new time. The third condition is that the initial
invariant polynomial (\ref{caz1}) have to generate the coupling
constant
\[K(\widetilde{L},\widetilde{L})=-b\,\]
instead of the Hamiltonian $\widetilde{H}$, as for the Kepler problem
\cite{ts98b}.

\begin{prop}
For each simple root $\b\in P$ and for any constant $b_\b\in{\mathbb
R}$ the following canonical transformation of the extended phase
space ${\cal M}_E$
\ben
d\widetilde{t}&=&e^{\b(q)}\cdot dt\,,\nn\\
 \label{dtodah}\\
\widetilde{H}_\b\,&=&e^{-\b(q)}\cdot
 \Bigl(\,H+b_\b\,\Bigr)\nn
\en
maps the Toda lattice into the other integrable system. This
canonical transformation induces the following transformation of the
Lax matrices
\bq
\widetilde{L}_\b=L-\widetilde{H}_\b\cdot \dfrac{e_\b}{a_\b}\,,
\qquad
\widetilde{L}_-={e^{-\b(q)}}\cdot L_-
\label{dtodal}
\eq
such that
\bq
\{\widetilde{H},\widetilde{L}\}=[\widetilde{L},\widetilde{L}_-]\,.
\label{dtodalax}
\eq
Here $H$, $L$ and $L_-$ are the Hamiltonian (\ref{todah}) and the Lax
matrices (\ref{todal},\ref{todalax}) for the corresponding Toda
lattice.

As for the Toda lattices, restrictions of the invariant function on
the orbit (\ref{dtodal}) give rise to the new set of integrals of
motion.
\end{prop}

There are $n={\rm rank}\,{\mathfrak g}$ functionally independent
invariant polynomials on ${\mathfrak g}$. Restricted to the orbit
${\cal O}_{ae}$ they remain functionally independent and give rise to
integral of motion for the Toda lattice. In (\ref{dtodal}) we
translate ${\cal O}_{ae}$ by adding to it a "constant" vector
proportional to the element of the universal enveloping algebra.  All
the invariant polynomials are invariant with respect to this
transformation. Thus, we can construct $n$ independent integrals of
motion in the involution for the system with the Hamiltonian
(\ref{dtodah}).

The number of the functional independent Hamilton functions
$\widetilde{H}_\b$, $\beta\in P$ depends on the symmetries of the
associated root system. For the closed Toda lattices associated with
the affine Lie algebras canonical time transformation has the similar
form. The associated with the Lax matrices $L$ (\ref{todal}) and
$\widetilde{L}$ (\ref{dtodal}) spectral curves depend on the choice
of a representation of $\mathfrak g$. Therefore, geometric
transformations of the spectral curves will be below considered on
the some examples only.

Of course, canonical transformations (\ref{dtodah}) induce
transformations of all the machinery developed for the Toda lattices.
As an example, for the Toda lattices associated to the classical
infinite series of the root systems $A_n,\,B_n,\,C_n$ and $D_n$ we
can introduce another $2\times 2$ Lax matrices \cite{ft87,kuzts89a}.
For brevity, here we restrict ourselves the $A_n$ root systems only.
In this case, the second Lax pair representation has the following
form
\[T=T_1\,T_2\cdots T_n\,,\qquad{\mbox{\rm where}}\qquad
T_j=\left(\begin{array}{cc}
  \l-p_j &a_j\, e^{q_j} \\
  -e^{-q_j}& 0
\end{array}\right)\,,\qquad a_j\in{\mathbb R}\,,\]
such that
\bq
\{H,T_j\}=T_j\,A_j-A_{j-1}\,T_j\,,\qquad
A_j=\left(\begin{array}{cc}
  \l &a_j\, e^{q_j} \\
  -e^{-q_j}& 0
\end{array}\right)\,.\label{2lax}
\eq
Change of the time (\ref{tr1}, \ref{dtodah}) associated with the root
$\b=\varepsilon_j-\varepsilon_{j+1}$ by
\[v(q)=e^{-\b(q)}=\exp(q_{j+1}-q_{j})\]
 gives rise to the following transformation of the Lax matrices
\bq
T=T_1\,T_2\cdots T_n\quad\mapsto\quad
\widetilde{T}=T_1\cdots T_{j-1}\cdot\left[T_j\,T_{j+1}+
\left(\begin{array}{cc}H+b & 0 \\  0 & 0\end{array}\right)\,\right]
\cdot T_{j+2}\cdots T_n\,,
\label{tr2l}
\eq
the Lax equations
\[
\{H,T\}=[T,A_n]\quad\mapsto\quad
\{\widetilde{H},\widetilde{T}\}=[\widetilde{T},v(q)\,A_n]\,,\]
and the spectral data of the Lax matrices
\ben
&\det T=\prod_{k=1}^n a_k\,, \qquad &{\rm tr}\,T=\l^n+\l^{n-1}\,P+
\l^{n-2}\,\left(\dfrac{P^2}2-H \right)+\ldots\,,\nn\\
&
\det\widetilde{T}=(a_j-\widetilde{H})\cdot\prod_{k\neq j}^n a_k
\,,\qquad
&{\rm tr}\,\widetilde{T}=\l^n+\l^{n-1}\,P+
\l^{n-2}\,\left(\dfrac{P^2}2+b\right)+\ldots\,.\nn
\en
All these transformations look like corresponding transformations for
the St\"ackel systems \cite{ts98b,ts98c}.

Let us consider some heuristic arguments concerning the Lax matrix
transformation (\ref{dtodalax}). Recall, constructions of the orbit
${\cal O}_{ae}$ and the corresponding classical $R$-matrix are
closely related to the principal grading of $\mathfrak g$. This
grading defines all the possible embedding of the three-dimensional
subalgebra $A_1\simeq sl(2)$ into ${\mathfrak g}$.

Let $\{{\ee,\ef,\eh}\}$ be generators of the Lie algebra $sl(2)$
\bq [{\eh,\ee}]={\ee}\,,\qquad
 [{\eh,\ef}]=-{\ef}\,,\qquad [{\ee,\ef}]=2{\eh}\,,\label{sl2}
\eq
and the element
\bq {\bf \Delta}={\eh}^2+\dfrac12({\ee\ef+\ef\ee})\,\label{caz}\eq
of the universal enveloping algebra be Laplace operator in $SL(2)$.
Let us consider infinite-dimensional irreducible representation $\cal
W$ of the Lie algebra $sl(2)$ in the linear space $V$ such that
\[{\cal W}:~\{{\ee,\ef,\eh}\}\to \{e,f,h\}\in {\rm End}(V)\,.\]
Let operator $e$ be invertible in $\rm{End}(V)$ and $\varphi(\Delta)$
be arbitrary function on the value of the Casimir operator
(\ref{caz}), then the mapping
\bq
e_-\to e'_-=e_-,\quad h\to h'=h,\quad e_+\to e_+'=e_++ e_-^{-1}\cdot
\varphi(\Delta)\,
\label{aut}
\eq
is an outer automorphism of the space of infinite-dimensional
representations of $sl(2)$ in $V$ \cite{ts98d}. These mappings shift
the spectrum of the Laplace operator $\bD$ (\ref{caz}) by the rule
\[ \Delta \to \Delta'=\Delta+\varphi(\Delta)\,.\]
In particular by $\varphi(\Delta)=-(\Delta+b)$ one gets
$\Delta\to\Delta'=-b$. So, instead of the spectrum of the Laplace
operator $\bf \Delta$ on the group $SL(2)$ one gets the spectrum of
the coupling constant $b$.

Now we turn to the Toda lattices related to the Lie algebra
${\mathfrak g}$, which contains subalgebras $A_1$ associated to the
roots $\b\in P$. The maps (\ref{dtodalax}) may be closely related to
the automorphism (\ref{aut}). Instead of the restriction $\Delta$
(\ref{caz}) of the Casimir operator $H(X)$ (\ref{caz1}) on $A_1$ one
have to substitute restriction $H(p,q)$ (\ref{todah}) of the same
Casimir operator on the orbit ${\cal O}_{a,e}$. Note, the outer
automorphism (\ref{aut}) non-trivial acts on the one nilpotent
subalgebra of $sl(2)$ only. Extension of this mapping on the orbit
${\cal O}_{ae}$ non-trivial acts on the constant one-point orbit of
${\mathfrak g}_-$, too.

Of course, the more justified consideration of the canonical
transformation of the time variable requires a more advanced
technique of representation theory.

\subsection{Examples}
Let us consider periodic three-particles Toda lattices associated to
the affine Lie algebra ${\cal L}(A_3)$. The corresponding Hamiltonian
$H$ (\ref{todah}) reads as
\bq
H=\dfrac12\,(p_1^2+p_2^2+p_3^2)+a_1\,e^{q_1-q_2}+a_2\,e^{q_2-q_3}+
a_3\,e^{q_3-q_1}\,,\label{a3h}
\eq
and the Lax matrix $L(\lambda,\mu)$ (\ref{todal}) has the following
form
\[
L(\lambda,\mu)=\left(\begin{array}{ccc}
  \l-p_1 & a_1\,e^{q_1-q_2} & {\mu}^{-1} \\
  1 & \l-p_2 & a_2\,e^{q_2-q_3} \\
  \mu\cdot a_3\,e^{q_3-q_1} & 1 & \l-p_3
\end{array}\right)\,.
\]
The Lax equations (\ref{todalax}) is linearized on the Jacobian of
following algebraic  hyperelliptic curve
\bq
{\cal C}:\quad {\rm det}\,L(\l,\m)=0\,,
\qquad
({a_1a_2a_3}{\mu}+\dfrac{1}{\mu})+\l^3+\l^2\,P+
\l\,\left(\dfrac{P^2}2-H\right)+K=0\,.
\label{todac}
\eq
Here complete momenta of the system $P=p_1+p_2+p_3$, Hamiltonian $H$
(\ref{a3h}) and additional integrals of motion $K$ are polynomials of
the first, second and third order in momenta, respectively.

The canonical transformation (\ref{dtodah}) maps the origin
Hamiltonian (\ref{a3h}) into the following Hamiltonian
\bq
\widetilde{H}={e^{q_2-q_1}}\cdot(H+b)\,,\qquad b\in{\mathbb
R}\,.\label{a3tr}
\eq
The associated Lax matrix $\widetilde{L}$ (\ref{dtodal}) is given by
\bq
\widetilde{L}=L-\left(\begin{array}{ccc}
  0 &0 & 0 \\
  a_1^{-1}\widetilde{H}& 0 & 0 \\
  0 & 0& 0
\end{array}\right)\,.
\eq
Its characteristic polynomial is
\bq
\widetilde{\cal C}:\quad {\rm det}\widetilde{L}(\l,\m)=0\,,\qquad
({a_1a_2a_3}{\mu}+\dfrac{a_1-\widetilde{H}}{a_1\,\mu})+\l^3+\l^2\,P+
\l\,\left(\dfrac{P^2}2+b\right)
+\widetilde{K}=0\,.
\eq
Here, in comparison with (\ref{todac}), the linear in $\l$ term is
proportional to the constant $b$ instead of the Hamilton function
$H$. The new Hamiltonian $\widetilde{H}$ is related to the second
spectral parameter now. For the first time similar transformations of
the Lax matrices and of their characteristic polynomials have been
observed in
\cite{ts98b,ts98c}.

The corresponding $2\times 2$ Lax matrix for the three-particle Toda
lattice is equal to
\[
T=T_1\,T_2\,T_3=
\left(\begin{array}{cc}
\l-p_1 & a_1\,e^{q_1} \\
-e^{-q_1} & 0\end{array}\right)\,
\left(\begin{array}{cc}
\l-p_2 & a_2\,e^{q_2} \\
-e^{-q_2} & 0\end{array}\right)\,
\left(\begin{array}{cc}
  \l-p_3 & a_3\,e^{q_3} \\
  -e^{-q_3} & 0\end{array}\right)\,.\]
Change of the time maps this matrix into the following matrix
\[\widetilde{T}=T_1\cdot
\left(\begin{array}{cc}
\l-p_2 & a_2\,e^{q_2} \\
-e^{-q_2}\,\left(1-a_1^{-1}\widetilde{H}\right)& 0\end{array}\right)
\cdot T_3\,,\]
such that
\ben
&{\det}\,T=a_1a_2a_3\,,\qquad &{\rm tr}\,T=\l^3+\l^2\,P+
\l\,\left(\dfrac{P^2}2-H \right)+K\,,\nn\\
\nn\\
&{\det}\,\widetilde{T}=(a_1-\widetilde{H}\,)\,a_2a_3\,,
\qquad &{\rm tr}\,\widetilde{T}=\l^3+\l^2\,P+
\l\,\left(\dfrac{P^2}2+b\right)+\widetilde{K}\,.\nn
\en
If we put $a_j=1$, the Poisson bracket relations for the initial
$2\times 2$ Lax matrices are closed into the Sklyanin quadratic
$r$-matrix algebra \cite{ft87,kuzts89a}
\[
\{\,{\on{T}{1}}(\l)\,,\,{\on{T}{2}}(\m)\,\}=
[\,r(\l-\m)\,,\,{\on{T}{1}}(\l)\,{\on{T}{2}}(\m)\,]\,,\qquad
r(\l-\mu)=\dfrac{\Pi}{\l-\mu}\,.
\]
Here the standard notations are introduced:
\[{\on{T}{1}}(\l)= T(\l)\otimes I\,,\qquad {\on{T}{2}}(\m)=I\otimes
T(\m)\,,
\]
and $\Pi$ is the permutation operator of auxiliary spaces
\cite{ft87}. Change of the time (\ref{a3tr}) maps the Sklyanin
$r$-matrix algebra into the following poly-linear algebra
\[
\{\,{\on{T}{1}}(\l)\,,\,{\on{T}{2}}(\m)\,\}=
[\,r(\l-\m)\,,\,{\on{T}{1}}(\l)\,{\on{T}{2}}(\m)\,]+
[\,r_{12}(\l,\m)\,,\,{\on{T}{1}}(\l)\,]+
[\,r_{21}(\l,\m)\,,\,{\on{T}{2}}(\m)\,]
\,,\]
with dynamical $r$-matrix
\ben
&&r_{12}(\l,\mu)=A_3(\l)\otimes\sigma\,T_3(\mu)\,,\qquad
\sigma=\left(\begin{array}{cc}
  1 & 0 \\
  0 & 0
\end{array}\right)\,,\nn\\
\nn\\
&&r_{21}(\l,\mu)=-\Pi\,r_{12}(\mu,\l)\,\Pi\,.\nn
\en
Here $T_3(\l)$ and $A_3(\l)$ were defined in (\ref{2lax}).

By using complete momenta $P$, we can reduce initial $6$-dimensional
phase space to the $4$-dimensional space. After this reduction and
after the following point canonical transformation of the coordinates
\ben
&&q_1=\dfrac12(1+i\sqrt{3})\ln{x}+\dfrac12(1-i\sqrt{3})\ln{y}\,,\nn\\
&&p_1=\dfrac12(1-\dfrac{i}{\sqrt{3}})x\,p_x+
\dfrac12(1+\dfrac{i}{\sqrt{3}})y\,p_y\,,\nn\\
\label{chang1}\\
&&q_2=\dfrac12(-1+i\sqrt{3})\ln{x}-\dfrac12(1+i\sqrt{3})\ln{y}\,,\nn\\
&&p_2=\dfrac12(-1-\dfrac{i}
{\sqrt{3}})x\,p_x-\dfrac12(1-\dfrac{i}{\sqrt{3}})y\,p_y\,,\nn
\en
the new Hamiltonian $\widetilde{H}$ reads as
\bq
\widetilde{H}=p_x\,p_y
+\dfrac{b}{xy}+a_2\,x^{z_1}\,y^{z_2}+a_3\,x^{z_2}\,y^{z_1}+a_1\,,
\label{da3h}
\eq
where $z_j$ are roots of the quadratic equation
$(z-z_1)(z-z_2)=z^2+3z+3=0$. For the first time the system defined by
$\widetilde{H}_1$ (\ref{da3h}) has been found in
\cite{dr35}.

In conclusion we present integrable systems related to the
two-particle Toda lattices associated to affine algebras $X^{(1)}_2$.
After an appropriate point transformation of coordinates, all the
Hamilton functions have a common form
\[\widetilde{H}=p_x\,p_y+\dfrac{b}{xy}+
a\,x^{z_1}\,y^{z_2}+c\,x^{s_1}\,y^{s_2}+d\,,\qquad a,b,c,d\in{\mathbb
R}
\]
where $z_{1,2}$ and $s_{1,2}$ be the roots of the different quadratic
equations. Below we show these equations only:
\ben
&A^{(1)}_3:\qquad &
\begin{array}{cc}
  z^2+3z+3=0 & s^2+3s+3=0 \\
\end{array}\nn\\
\nn\\
&B^{(1)}_2~C^{(1)}_2:\qquad &
\begin{array}{cc}
  z^2+4z+5=0 & s^2+4s+5=0 \\
  z^2+2z+2=0 & s^2+3s+5/2=0
\end{array}\nn\\
\nn\\
&D^{(1)}_2:\qquad &
\begin{array}{cc}
  z^2+2z+2=0 & s^2+2s+2=0 \\
  z^2+2z+2=0 & (s+2)^2=0
\end{array}\nn\\
\nn\\
&G^{(1)}_2:\qquad &
\begin{array}{cc}
  z^2+2z+4=0 & s^2+5s+7=0 \\
  z^2+2z+4=0 & s^2+3s+3=0 \\
  z^2+3z+7/3=0 & s^2+3s+3=0
\end{array}\nn
\en
The corresponding second integrals of motion $K$ are polynomials
third, fourth and sixth order in momenta. Note, for the algebra
$A^{(1)}_3$ all the three Hamiltonians $H_\b,~\b\in P$ are
equivalent. Two different Hamilton function (\ref{dtodah}) associated
with the algebras $B^{(1)}_2$, $C^{(1)}_2$ and $D^{(1)}_2$. For the
$G^{(1)}_2$ algebra we have three different Hamiltonians
(\ref{dtodah}).

\section{The St\"{a}ckel systems}
\setcounter{equation}{0}
In this Section we propose an interesting extension of the integrable
family of the St{\"a}ckel systems \cite{st95}. As an example, we
discuss here a family of the two-dimensional integrable systems in
detail. In the two limiting cases, the corresponding systems possess
the following Hamilton functions
\ben
H&=&p_x^k\,p_y^k+ \a\,\bigl(x\,y\bigr)^{-
\textstyle{{k}\over{k+1}} }\,,\qquad
\a\,,k\in{\mathbb R}\,,\nn\\
\label{gfk}\\
H&=&p_x^k+p_y^k+ \a\,\bigl(x\,y\bigr)^{-
\textstyle{{k}\over{k+1}} }\,,\nn
\en
where $\a$ and $k$ are arbitrary parameters. At $k=1$ the first
Hamiltonian coincides with the Hamiltonian of the Kepler problem. At
$k=2$ the second integrable Hamiltonian has been found by Fokas and
Lagerstrom  \cite{fl80}. It is known, both these systems are dual to
the some St\"{a}ckel systems (see review \cite{hgdr84} and references
therein).  By using similar duality we shall prove integrability of
the general systems (\ref{gfk}).

Let variables $\{p_j,q_j\}_{j=1}^n$  be coordinates in the phase
space ${\mathbb R}^{2n}$ with the standard Poisson brackets
\[\{p_j,q_k\}=\d_{jk}\,,\qquad j,k=1,\ldots,n\,.\]
Let us consider two integrable hamiltonian systems on the common
phase space ${\mathbb R}^{2n}$. These systems are defined by the two
sets of independent integrals of motion $\{I_j\}_{j=1}^n$ and
$\{J_j\}_{j=1}^n$, in the involution
\[ \{I_j,I_k\}=0\,\qquad {\rm and} \qquad
\{J_j,J_k\}=0\,,\quad j,k=1,\ldots,n\,.\]
By the Liouville theorem for given integrable systems we can
introduce two family of the action-angle variables
$\{s_j,\varphi_j\}_{j=1}^n$ and
$\{\tilde{s}_j,\tilde{\varphi}_j\}_{j=1}^n$ for the each of
integrable systems. Integrals of motion depend on the action
variables only
\[I_k=I_k(s_1,\ldots,s_n)\,,
\qquad J_k=J_k(\tilde{s}_1,\ldots,\tilde{s}_n)\,.\]
According to expansion (\ref{exph}), if two a'priori different
systems of the action coordinates are related
\bq
\tilde{s}_j=\tilde{s}_j\,(s_1,\ldots,s_n)
\label{ogr1}
\eq
then the canonical transformations of the extended phase space at
\[ v(p,q)=v(s)=I_k(p,q)\,,\qquad \mbox{\rm or}\qquad
v(p,q)=v(s)=J_k(p,q)\] map a given pair of integrable system into the
third integrable system.

Thus, we have to verify condition (\ref{ogr1}) and have to construct
explicitly the third family of integrals of motion in the initial
physical variables. For instance, by using inner product of the two
independent vectors of integrals ${\bf I}$ and ${\bf J}$ in ${\mathbb
R}^n$ we can introduce some antisymmetric matrix ${\cal K} =({\bf
I}\otimes {\bf J})$. Any column or row of this matrix defines a set
of the $n-1$ independent functions
\[{\cal K}_{ij}=({\bf I}\otimes {\bf J})_{ij}
=I_i\,J_j-I_j\,J_i\,,\qquad i,j=1,\ldots,n\,.\] Under some
restriction on the initial integrals $\{I_j\}$ and $\{J_j\}$ we could
construct from them a third integrable system on the same phase
space.
\begin{prop}
If all the differences of integrals of motion $(I_j-J_j)$ with the
common index $j=1,\ldots,n$ are in the involution
\bq
\Bigl\{I_j-J_j\,,I_k-J_k\Bigr\}=0\,,\qquad j,k=1,\ldots,n\,,\label{usl}
\eq
then the ratio of integrals
\bq
K_m=\dfrac{I_m}{J_m}
\label{newh}
\eq
and $n-1$ functions $K_j$, $j\neq m$
\bq K_j= \dfrac{{\cal K}_{mj}}{J_m}=\dfrac{I_m\,J_j-I_j\,J_m}{J_m}=
 K_m\,J_j-I_j\,,\qquad
m\neq j=1,\ldots,n
\label{newi}
\eq
are integrals of motion for new integrable system on the same phase
space.
\end{prop}
By definition all the new integrals $K_m$ and $K_j$ are functionally
independent. Thus, the proof is straightforward calculation of the
following Poisson brackets
\[
\{K_m,K_j\}=\dfrac{I_m}{J_m1^2}\,\Bigl(\,\{I_m,J_j\}
+\{J_m,I_j\}\,\Bigr)=-\dfrac{I_m}{J_m^2}\,\Bigl\{I_m-J_m,I_j-J_j\Bigr\}=0\,\,.
\]
and
\ben
\{K_j,K_k\}&=&K_m\,\{J_j,K_k\}-\{I_j,K_k\}=\nn\\
&=&J_k\,\Bigl(K_m\,\{J_j,K_m\}-\{I_j,K_m\}\Bigr)-
K_m\,\Bigl(\{J_j,I_k\}+\{I_j,J_k\}\Bigr) =\nn\\
&=&J_k\,(\{K_j,K_m\}+K_m\,\Bigl\{I_j-J_j,I_k-J_k\Bigr\}=0\,.
\qquad j\neq k\neq m
\nn
\en
Thus, canonical transformation (\ref{newh}-\ref{newi}) of the
extended phase space
\[\bigr({\bf I},{\bf J}\bigl)\to {\bf K}\]
preserves the property of integrability. To apply this transformation
we have to find two integrable systems satisfying condition
(\ref{usl}). Below we prove that the St\"{a}ckel integrable systems
\cite{st95}  may be considered as the main example of the systems satisfying condition (\ref{usl}).

Let us briefly recall some necessary facts about the St\"{a}ckel
systems
\cite{st95,ts97d}. The nondegenerate $n\times n$ St\"{a}ckel matrix $\bs$, whose $j$ column $\sm_{kj}$ depends only on $q_j$
\[\det \bs\neq 0\,,\qquad \dfrac{\partial \sm_{kj}}{\partial q_m}=0\,,
\quad j\neq m\]
defines functionally independent integrals of motion
$\{I_k\}_{k=1}^n$
\bq
I_k=\sum_{j=1}^n c_{jk}\left(p_j^2+U_j\,(q_j)\,\right)\,,
\qquad c_{jk}=\dfrac{\bs_{kj}}{\det\bs}\,,
\label{int1}
\eq
which are quadratic in momenta. Here ${\bf C}=[c_{ik}]$ denotes
inverse matrix to $\bs$ and  $\bs_{kj}$ be cofactor of the element
$\sm_{kj}$.

We can see that in practical circumstances the St\"{a}ckel approach
is not very useful because it is usually unknown what canonical
transformation have to be used to transform a Hamiltonian
(\ref{int1}) to the natural form
$H=T(p_1,\ldots,p_n)+V(q_1,\ldots,q_n)$. This problem was partially
solved for the St\"{a}ckel systems with a common potential
$U_j=U,~j=1,\ldots,n$ only
\cite{ts97d}.

According to \cite{ts98b}, if the two St\"{a}ckel matrices $\bs$ and
$\tilde{\bs}$ be distinguished the $m$-th row only
\[\sm_{kj}=\tilde{\sm}_{kj}\,,\qquad k\neq m\,,\]
the corresponding St\"{a}ckel systems with a common set of potentials
$U_j$ and with the Hamilton functions $I_m$ and $\widetilde{I}_m$ are
related by canonical change of the time
\bq
\widetilde{I}_m \longleftrightarrow {I}_m\,,\qquad \widetilde{I}_m=
\dfrac{I_m(p,q)}{v(q)}\,. \label{ntrans1}
\eq
where
\[v(q_1,\ldots,q_n)=
\dfrac{\det\widetilde{\bs}(q_1,\ldots,q_n)}{\det{\bs}(q_1,\ldots,q_n)}\]
The canonical transformation (\ref{ntrans1}) connects two St\"{a}ckel
systems with the different St\"{a}ckel matrices and with the common
set of potentials $U_j$.

Let us consider a pair of the St\"{a}ckel systems with a common
St\"{a}ckel matrix $\bs$ and with the different potentials. Namely,
in addition to the system with integrals $\{I_k\}$ (\ref{int1}), we
introduce the second integrable system with the similar integrals of
motion
\bq
J_k=\sum_{j=1}^n c_{jk}\left(p_j^2+W_j\,(q_j)\,\right)\,,\quad
k=1,\ldots,n\,.
\label{int2}
\eq
Here even one potential $U_j\,(q_j)$ has to be functionally
independent on the corresponding potential $W_j\,(q_j)$.

\begin{prop}
Any two integrable systems defined by the same St\"{a}ckel matrix
$\bs$ and by the functionally independent potentials $U_j\,(q_j)$ and
$W_j\,(q_j)$ satisfy the necessary condition (\ref{usl}) of the
previous proposition. Thus, the ratio of the two St\"{a}ckel
integrable Hamiltonians defines new integrable system
\bq
K_m \longleftrightarrow (I_m,J_m)\,,\qquad
K_m=\dfrac{I_m}{J_m}\,.\label{ntrans2}
\eq
\end{prop}
It is obvious, all the integrals $I_k$ and $J_k$ are differed by the
potential part
\[(I_k-J_k)=\sum_{j=1}^2 c_{jk}\left[\,U_j(q_j)-W_j(q_j)\,\right]\]
depending on coordinates  $\{q_j\}$ only.  Thus, systems with a
common St\"{a}ckel matrix $\bs$ satisfy condition (\ref{usl}).

The Hamilton function (\ref{ntrans2}) has the following form
\bq
H=K_m=\dfrac{ \sum_{j=1}^n c_{jm}\left[p_j^2+U_j\,(q_j)\right]
}{\sum_{j=1}^n c_{jm}\left[p_j^2+W_j\,(q_j)\right] }\,.
\label{newhs}
\eq
This Hamiltonian $H$ is a rational function in momenta, but next one
can try to use canonical transformations to simplify it. In rare
case, one obtains again a natural type Hamilton function as will be
shown below.

As for the usual St\"{a}ckel system, the common level surface of the
new integrals (\ref{newh})
\[
M_\a=\left\{z\in {\mathbb R}^{2n}:~K_j(z)=\a_j\,,~j=1,\ldots,n
\right\}
\]
is diffeomorphic to the real torus. Namely, substituting
\[V_j\,(q_j)=(1-\a_m)^{-1} \bigl(\,U_j\,(q_j)-\a_m\,W_j\,(q_j)\,\bigr)\,,\]
into the definitions (\ref{newhs}) and (\ref{int1},\ref{int2}) we
obtain the following equations
\ben
\sum_{j=1}^2
c_{jm}\left[\,p_j^2+V_j\,(q_j)\right]&=&0=\b_m\,,\nn\\
\label{abt}\\
\sum_{j=1}^2
c_{jk}\left[\,p_j^2+V_j\,(q_j)\right]&=&-\dfrac{\a_k}{1-\a_m}=\b_k\,.
\nn
\en
After multiplication of these equations by the St\"{a}ckel matrix one
immediately gets
\[
p_j^2=\left(\dfrac{\partial {\cal S}}{\partial q_j}\right)^2=
\sum_{k=1}^n \b_k\sm_{kj}(q_j)-V_j(q_j)\,,
\]
where ${\cal S}(q_1,\ldots,q_n)$ is a reduced action function  The
corresponding Hamilton-Jacobi equation on $M_\a$
\[
\dfrac{\partial {\cal S}}{\partial t}+
H(t,\dfrac{\partial {\cal S}}{\partial q_1},\ldots,
\dfrac{\partial {\cal S}}{\partial q_n},q_1,\ldots,q_n)=0\,,
\qquad\Rightarrow\qquad
c_{jm}\,\dfrac{\partial {\cal S}}{\partial q_j}\,
\dfrac{\partial {\cal S}}{\partial q_j}=E\,,
\]
admits the variable separation ${\cal S}(q_1,\ldots,q_n)=\sum_{j=1}^n
{\cal S}_j(q_j)$, where
\[\qquad {\cal S}_j(q_j)=\int{
\sqrt{\sum_{k=1}^n
\b_k\sm_{kj}(q_j)-V_j(q_j)~~}
~d q_j}\,.
\]
Thus, coordinates $q_j\,(t,\a_1,\ldots,\a_n)$ are determined from the
equations
\[
\sum_{j=1}^n\int\dfrac{\sm_{kj}(\l) d\l}
{\sqrt{\sum_{k=1}^n \b_k\sm_{kj}(\l)-V_j(\l)}}=\d_k\,,
\qquad k=1,\ldots,n\,,
\]
where the constants $\a_j$ and $\b_j$ are related by (\ref{abt}).

Thus, the solution of the problem is reduced to solving a sequence of
one-dimensional problems, which is the essence of the method of
separation of variables. Next, the integration problem for equation
of motion is reduced to solution of the inverse Jacobi problem in
framework of the algebraic geometry \cite{ts98b}.

Recall, the first canonical transformation (\ref{ntrans1}) may be
related to the ambiguity of the Abel map. So, it would be interesting
to investigate the underlying algebro-geometric origin of the
construction (\ref{ntrans2}).

For the some St\"ackel systems with uniform potentials
$U_j=U\,,~j=1,\ldots,n$ we can construct the $2\times 2$ Lax matrices
\cite{ts97d}. Of course, transformation of the time (\ref{ntrans1}) induces transformation of the Lax matrices. As for the Toda lattices, these transformations may be reduced to the adding a constant with respect to the new time matrix \cite{ts98b,ts98c} to the initial Lax matrix. Transformations of the Lax matrices by new mapping of the time (\ref{ntrans2}) will be published elsewhere.

\section{Examples}
As we have mentioned before, the Hamiltonian $H$ (\ref{newhs}) have a
rather unusual expression. However, in some cases suitable canonical
transformations can reduce it to a sum of the kinetic energy and the
potential energy. Note, necessary choice of such transformations is a
generic problem for all the St\"{a}ckel systems. Thus, in this
Section we present several concrete two-dimensional systems only.

Let us consider polar coordinate system on plane with the usual
coordinates $(p_r,r)$ and $(p_\phi,\phi)$ instead of
$(p_{1,2},q_{1,2})$, respectively. We take the first system with the
axially symmetric potential
\bq
I_1=p_r^2+\dfrac{p_\phi^2}{r^2}-a^2\,r^{2k}+b\,,\qquad
I_2=p_\phi\,,\qquad a,\,b,\,n \in{\mathbb R}\,.
\label{sys1}
\eq
The second system is associated to a free motion
\bq
J_1=p_r^2+\dfrac{p_\phi^2}{r^2}\,,\qquad J_2=p_\phi\,.
\label{sys2}
\eq
These systems belong to the St\"{a}ckel family of integrable system
with the following common St\"{a}ckel matrix
\[\bs=\left(\begin{array}{cc} 1& 0\\  -r^{-2}&1\end{array}\right)
\]
In this case $\sm_{12}=c_{21}=0$ and $I_2-J_2=0$, it allows us to use
the second integrals in the non-St\"{a}ckel form (\ref{int1}).

The ratio (\ref{newh}) of the Hamiltonians (\ref{sys1}-\ref{sys2})
may be rewritten in the form (\ref{gfk}) by using the set of
canonical transformations. Let us begin with the usual transformation
of the curvilinear coordinates to the cartesian coordinates
\ben
&r=\sqrt{u\,v~}\,,\qquad
&\phi=i\arctan\left(\dfrac{u-v}{u+v}\right)\,,\nn\\
\label{trans1}\\
&p_r=-\dfrac{u\,p_u+v\,p_v}{r}\,,
\qquad
&p_\phi=i\,\left(u\,p_u-v\,p_v\right)\,.\nn
\en
Then, one permute coordinates and momenta  $(u\leftrightarrow p_u)$
and $(v\leftrightarrow p_v)$ such that new Hamiltonian (\ref{newhs})
becomes polynomial in momenta
\bq
H=\dfrac{I_1}{J_1}=-\dfrac{a^2}4\,\dfrac{\,p_u^k\,p_v^k\,}
{\,u\,v}+\dfrac{b}{4\,u\,v}+1\,.
\label{h31}
\eq
In conclusion, we have to use the point canonical transformation
\ben
&&p_x=p_u\,u^{\textstyle{{1}\over{k+1}}}\,,\qquad
    x=(1+1/k)~u^{\textstyle{{k}\over{k+1}}} \,,\nn\\
    \label{qtrans}
&&p_y=p_v\,v^{\textstyle{{1}\over{k+1}}}\,,\qquad
    y=(1+1/k)~v^{\textstyle{ {k}\over{k+1}}} \,,\nn
\en
which converts the Hamiltonian (\ref{h31}) into the following form
\[H=p_x^k\,p_y^k+\a\,\bigl(x\,y\bigr)^{-
\textstyle{{k}\over{k+1}} }+\b\,,\]
after multiplication on a suitable constant and rescaling parameters.

Note, the St\"{a}ckel matrix $\bf S$ and the set of potentials
$U_j(q_j)$ are determined on the half of the phase space ${\mathbb
R}^{2n}$ and depend on coordinates $q_j$ only. Thus, we have some
freedom related to the different canonical transformations of momenta
\ben
&&(\,p_1,\ldots\,p_n\,)\to(\,\widetilde{p}_1,\ldots,\widetilde{p}_n\,)\,,\nn\\
\label{trans}\\
&&p_i-\widetilde{p_i}=2\,\dfrac{\partial F(q_1,\ldots\,q_n)}{\partial
q_i}
\,,\qquad p_i+\widetilde{p}_i=0\,.\nn
\en
Here $ F(q_1,\ldots\,q_n)$ is a generating function of the
transformations (\ref{trans}) depending on coordinates $\{q_j\}$,
which are invariant with respect to transformation (\ref{trans}).

If condition (\ref{usl}) is invariant under these canonical
transformations, we can apply them (\ref{trans}) to construct new
integrable systems by the rule (\ref{newh}). Although no general
procedure exists for this, one interesting case is known.

As above, one takes system with the axially symmetric potential
(\ref{sys1}) and system associated with a free motion with integrals
\bq
J_1=\tilde{p}_r^2+\dfrac{\tilde{p}_\phi^2}{r^2}\,,\qquad
J_2=\tilde{p}_\phi\,.
\label{sys3}
\eq
New momenta $(\tilde{p}_r\,,\tilde{p}_\phi\,)$ relate with old ones
$(p_r\,,p_\phi\,)$ by canonical transformation (\ref{trans})
\bq
\tilde{p}_r=p_r-
\dfrac{\partial\,f(r)}{\partial r}\,\dfrac{\cos(n\,\phi)}{n}\,,\qquad
\tilde{p}_\phi=p_\phi+f(r)\,\sin(n\,\phi)\,.
\label{rphi}
\eq
Here $f(r)$ be any function on variable $r$ and $n$ be arbitrary
parameters.

Both these systems belong to the St\"{a}ckel family of integrable
system associated with a common St\"{a}ckel matrix. In this case
$I_2-J_2\neq 0$ and condition (\ref{usl}) do not invariant by
transformation (\ref{rphi}).  Let the second integrals be square root
from the usual St\"{a}ckel integrals (\ref{int1}). If this form of
the second integrals are used, the pair of the systems (\ref{sys1})
and (\ref{sys3}) satisfies condition (\ref{usl}) by
\[\dfrac{f'(r)}{f(r)}=\dfrac{n}{r}~\Rightarrow~f(r)=c\,r^n\,,
\qquad c\in{\mathbb R}\,. \]
Let us get over to the cartesian coordinates (\ref{trans1}) and
conjugated momenta
\bq
\tilde{p}_r=-\dfrac{u\,p_u+v\,p_v}{r}\,,
\qquad
\tilde{p}_\phi=i\,\left(u\,p_u-v\,p_v\right)\,.
\eq
After permutation coordinates and momenta $(u\leftrightarrow p_u)$
and $(v\leftrightarrow p_v)$ (analog of the Fourier transformation in
the quantum mechanics), one gets
\bq
H=\dfrac{I_1}{J_1}=
\dfrac{1}{4\,u\,v}\,\bigl(c^2\,p_u^{n-1}\,p_v^{n-1}-
a^2\,p_u^{k}\,p_v^k-2\,c\,(v\,p_u^{n-1}\,+u\,p_v^{n-1}\,)\bigr)
+\dfrac{b}{4\,u\,v}+1\,.\label{h32}
\eq
At $c=0$ we discuss this system before (\ref{h31}). Now, we consider
the second limiting case by
\[a=c\,,\qquad k=n-1\,,\]
that simplifies potential part of the Hamiltonian (\ref{h32}).

As above, the  point canonical transformation (\ref{qtrans}) converts
the Hamiltonian (\ref{h32}) into the following form
\[H=p_x^k+p_y^k+\a\,\bigl(x\,y\bigr)^{-
\textstyle{{k}\over{k+1}} }+\b\,,\]
after multiplying on a suitable constant and rescaling parameters.

Thus, we present a family of two-dimensional integrable systems,
which includes the Kepler and Fokas-Lagerstrom potentials
simultaneously.

\section{Acknowledgement}
Author is very grateful to I.V.Komarov whose sincere attention helped
me to continue this work. This work was partially supported by RFBR
grant 99-01-00698.
% ------------------------------------------------------------------------
%\bibliography{qism}
%\bibliographystyle{plain}

% ------------------------------------------------------------------------
\end{document}